\documentclass{emulateapj}

\usepackage{float}
\usepackage{amsmath}
\usepackage{epsfig,floatflt}

\usepackage{graphicx}
\usepackage[]{subfigure}
\usepackage{color}

\begin{document}

 \title{Non-Gaussianities in the local curvature of the 5-year WMAP data}

\author{\O ystein Rudjord\altaffilmark{1,2}}
\email{oystein.rudjord@astro.uio.no}

\author{Nicolaas E. Groeneboom\altaffilmark{1,2}}

\author{Frode K. Hansen\altaffilmark{1,2}}

\author{Paolo Cabella\altaffilmark{3}}

\altaffiltext{1}{Institute of Theoretical Astrophysics, University of
  Oslo, P.O.\ Box 1029 Blindern, N-0315 Oslo, Norway}

\altaffiltext{2}{Centre of Mathematics for Applications, University of
  Oslo, P.O.\ Box 1053 Blindern, N-0316 Oslo, Norway}

\altaffiltext{3}{Dipartimento di Fisica, Universita di Roma 'Tor Vergata', Via della Ricerca Scientifica 1, I-00133, Roma, Italy}

\date{\today}

\begin{abstract}
  Using the 5 year WMAP data, we re-investigate claims of
  non-Gaussianities and asymmetries detected in local curvature
  statistics of the 1 year WMAP data. In Hansen et al 2004, it was
  found that the northern ecliptic hemisphere was non-Gaussian at the
  $\sim1\%$ level testing the densities of hill-, lake and saddle
  points based on the second derivatives of the CMB temperature
  map. The 5 year WMAP data has a much lower noise level and better
  control of systematics. Using these, we find that the anomalies are
  still present at a consistent level. Also the direction of maximum
  non-Gaussianity remains. Due to limited availability of computer
  resources, Hansen et al. 2004 were unable to calculate the full
  covariance matrix for the $\chi^2$ test used. Here we apply the full
  covariance matrix instead of the diagonal approximation and find
  that the non-Gaussianities disappear and there is no preferred
  non-Gaussian direction. We compare with simulations of weak lensing
  to see if this may cause the observed non-Gaussianity when using
  diagonal covariance matrix. We conclude that weak lensing does not
  produce non-Gaussianity in the local curvature statistics at the
  scales investigated in this paper. The cause of the non-Gaussian
  detection in the case of a diagonal matrix remains unclear.
\end{abstract}

\maketitle

\section{Introduction}
\label{sec:introduction}
During recent years, the investigations of the cosmic microwave
background (CMB) has proved to be the most compelling addition to our
understanding of the early universe. Observations of the CMB
anisotropies, like those obtained by the Wilkinson Microwave
Anisotropy Probe (WMAP) experiment \citep{bennett:2003, hinshaw:2007},
have provided us with profound insight on the composition of structure
in our universe. Combined with previous experimental knowledge and a
sound theoretical framework, the concordance model of $\Lambda$CDM has
been established.

The $\Lambda$CDM model relies on the framework of inflation. Inflation
was initially proposed as a solution to the horizon and flatness
problem \citep{guth:1981}. Additionally, it established a highly
successful scenario for the formation of primordial density
perturbations, providing the required seeds for the large-scale
structures in the universe.  Eventually, these later gave rise to the
temperature anisotropies in the cosmic microwave background radiation
that we observe today. It is assumed that Gaussian fluctuations in the
vacuum field have given rise to these perturbations, so the
fluctuations in the CMB map we observe today are also expected to be
near Gaussian.

Inflation predicts that the observed universe should be statistically
isotropic on large scales. However, during recent years, various
anomalies that contradict statistical isotropy have been discovered in
the WMAP data \citep{de Oliveira-Costa:2004, eriksen:2004a,
  hansen:2009,hoftuft,eriksen:2004b,hansen:2004a,tegmark2003,
  groeneboom:2008b, groeneboom:2009a, vielva:2004}.  If these effects
are confirmed by the data from the Planck experiment, various
anisotropic universe models should be seriously considered.

In order to investigate the properties of different inflationary
models, it is important not only to pursue deviations from statistical
isotropy, but also any deviations from Gaussianity.  Single-field
inflation models usually predict a CMB statistically close to
Gaussian, but more exotic models may give rise to a larger
contribution of non-Gaussianity. In this paper, we focus on a method
for testing general deviations from Gaussianity that are not obviously
connected to inflationary non-Gaussianity. However, some effort has
been made to use the local curvature to estimate the non-Gaussianity
parameter $f_{NL}$ \citep{cabella2005}.

\cite{dore:2003} presented a framework for investigating
non-Gaussianities based on the properties of the local curvature of
the CMB temperature map.  By calculating the second-order derivatives,
it is possible to classify each pixel as a ``hill'', ``lake'' or
``saddle'' based on the eigenvalues of the Hessian matrix. If the map
is first smoothed with a Gaussian beam, it is possible to extract the
local curvature properties on scales given by the FWHM of the beam. A
temperature threshold $T_{\textrm{t}}$ is introduced, where CMB
temperature values below a provided threshold are ignored.  Starting
with a negative temperature threshold, the fraction of hill, lake and
saddle points that have temperatures above the threshold are
counted. By performing this analysis on simulated isotropic Gaussian
maps, it is possible to estimate what the fraction of hills, lakes and
saddle points should be for increasing $T_{\textrm{t}}$. This graph is
then compared with a similar examination of experimental data using a
$\chi^2$ test.  It is is also possible to predict this plot
theoretically, as done by \cite{dore:2003}. However, several issues
will complicate the task. First, one needs to apply a galaxy mask that
effectively removes about $20 \%$ of the data. There are also point
sources to be considered, along with the possibility of excluding
large chunks of the sky in order to focus the analysis on specific
areas.

\cite{hansen:2004b} applied their framework on the 1-yr WMAP data,
where no statistical deviation from Gaussianity was found on a
full-sky coverage. However, when the authors analyzed independent
hemispheres on the CMB sky, they found a non-Gaussian signature on
hemispheres centered near the ecliptic poles. The authors therefore
suggested that the effect might be due to systematics, and that the
direction seems consistent with the results by
\cite{eriksen:2004a}. However, due to limited computer resources, the
authors built their statistics using only 128 and 512 Gaussian
simulations, which may be too few for obtaining convergence. For the
same reason, they were unable to obtain a full coverged covariance
matrix for the $\chi^2$ analysis and they therefore ignored the
correlations between threshold levels. In addition, the 1-year WMAP
data are much more noisy compared to the 5-year data. In this paper,
we re-analyze the 5-year WMAP data using an independent and more
robust code with a greatly increased number of Gaussian simulations.
We also test whether weak lensing may contribute to a non-Gaussian
signal in the local curvature statistics.

\section{Method}
\begin{figure}
\mbox{\epsfig{file=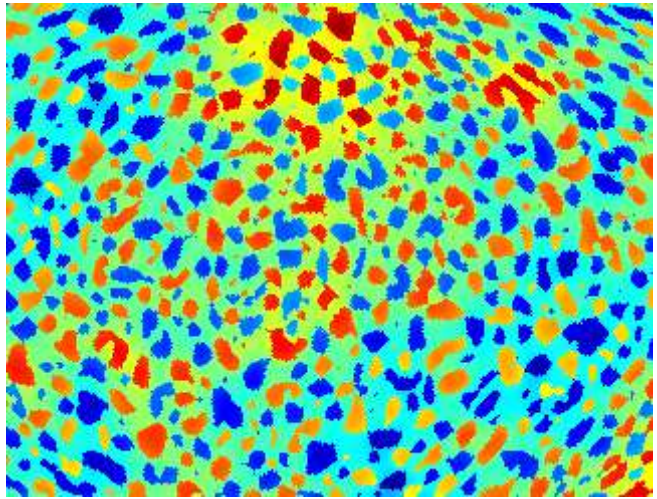, width=\linewidth,clip=}}
\caption{The classification of hills (red parts), lakes (blue parts) and saddles (neutral, in between). The original CMB map is added in the background for illustrative purposes. The beam FWHM used in this example is 500' on an isotropic realization of the best-fit $\Lambda$CDM power spectrum.}
\label{fig:classification}
\end{figure}

The methods in this paper are based on the work done by
\cite{hansen:2004b} and
\cite{dore:2003}. We are interested in studying the curvature
properties of the temperature fluctuations in the CMB map. Every
point on a two-dimensional surface embedded in 3-space can be
characterized as being a hill, lake or saddle. In the case of
the CMB, we usually express coordinates on $\mathbb S^2$ in spherical
coordinates $(\theta, \phi)$.  A hill/lake is where the curvature is
negative/positive in both the $\theta$ and $\phi$ directions, whereas
a saddle point has negative curvature for one coordinate and
positive for the other. 
While increasing the threshold value $T_t$, we count the fraction
of hill, lake and saddle points for all the pixels with temperature larger than $T_t$. A simulated CMB map
with the classified curvature areas added on top for illustrative purposes is depicted in Figure
\ref{fig:classification}.

\subsection{The local curvature}
\label{sec:curvature}
The CMB map is first smoothed with a beam corresponding to the scales
that we are interested in. We then normalize the temperature map
$T$ with its standard deviation and calculate its various first and
second-order covariant derivatives given the spherical coordinates
$\theta$ and $\phi$. That is, we obtain $\frac{dT}{d\theta}$,
$\frac{dT}{d\phi}$, $\frac{d^2T}{d\theta^2}$, $\frac{d^2T}{d\phi^2}$
and $\frac{d^2T}{d\theta d\phi}$. For each pixel in the the normalized
temperature map $T$, we calculate the Hessian matrix 
and calculate its eigenvalues $\lambda_1$ and $\lambda_2$ (as described in \cite{monteserin:2005}). Then,
depending on the sign of the eigenvalues, we classify every point on
the sphere as a hill ($\lambda_1 > 0$, $\lambda_2 > 0$), lake
($\lambda_1 < 0 $, $\lambda_2 < 0$) or saddle point
($\lambda_1,\lambda_2 < 0$).

\begin{figure}
\mbox{\epsfig{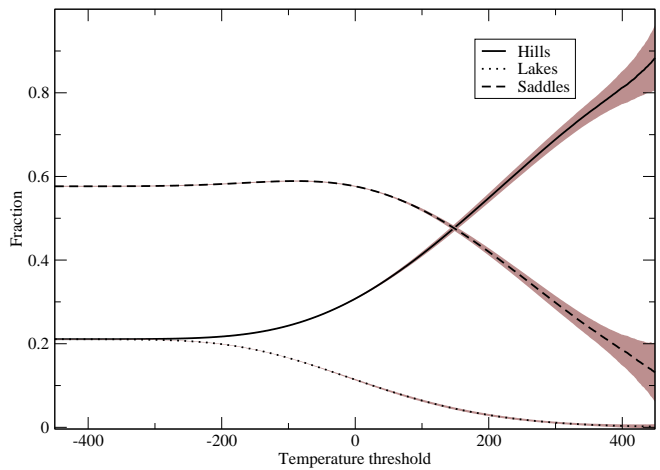}}
\caption{
The fraction of hills, lakes and saddle pixels in a simulated Gaussian map with $n_{\textrm{side}}=512$. The simulation is generated from the best-fit powerspectrum from the WMAP 5 year release. The
black line depicts the mean, while the shaded area is the $68 \%$ confidence area.}
\label{fig:hills_lakes_saddles}
\end{figure}

By performing this analysis on
simulated isotropic Gaussian maps, it is possible to estimate what the
mean fraction of hills, lakes and saddle points should be for
increasing $T_{\textrm{t}}$, together with the standard
deviation. An example of such a graph is presented in Figure \ref{fig:hills_lakes_saddles}.

\subsection{The extended mask}
\label{sec:extended_mask}
When working with experimental CMB data, we need to remove a portion of the
sky due to foreground residuals. 
In this paper, we operate with several masks, most
notably the WMAP KQ85 sky cut \citep{gold:2008}, which removes 18\% of
the sky. The edges of the mask need special consideration, due to the
differentiation of the map. We therefore perform the following procedure for generating suitable CMB masks:
\begin{itemize}
  \item[1.] Start with the 5-year WMAP KQ85 or KQ75  mask without point source holes.
  \item[2.] The mask is expanded by $b^\circ$, where $b$ is the smoothing angle FWHM to avoid effects of the smoothed mask. Note that this step was not used in \cite{hansen:2004b}. We thus use an extended mask which is considerably much larger than the one used in \cite{hansen:2004b}. As pointed out below, we have compared results also with the smaller extended mask for which this step was omitted. 
  \item[3.] Point source holes are added to the mask after expansion. If added before smoothing, the smeared point sources would nearly fill the mask.
  \item[4.] The mask is differentiated, and then normalized. By normalized, we mean that each pixel $p_i$ is set to $p_i = \frac{|p_i|}{p_{max}}$.
  \item[5.] If a pixel in a normalized, differentiated map has value above a certain threshold, the pixel is masked out. We have used the threshold $0.02$ throughout this paper.
\end{itemize}

\begin{figure}
\mbox{\epsfig{file=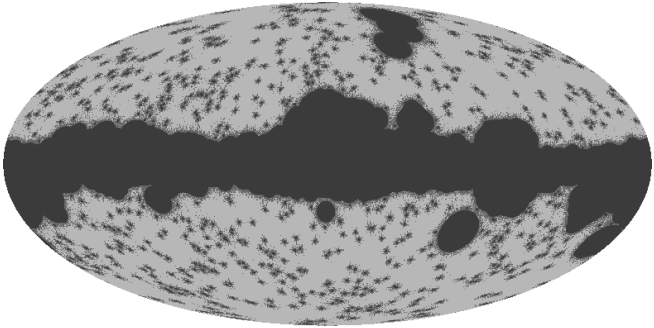, width=\linewidth,clip=}}
\caption{The KQ85 mask expanded with $5^\circ$ and point-sources
  added. The expansion from differentiation is indicated in grey.}
\label{fig:mask1}
\end{figure}

\subsection{$\chi^2$ statistics}
\begin{figure}
\mbox{\epsfig{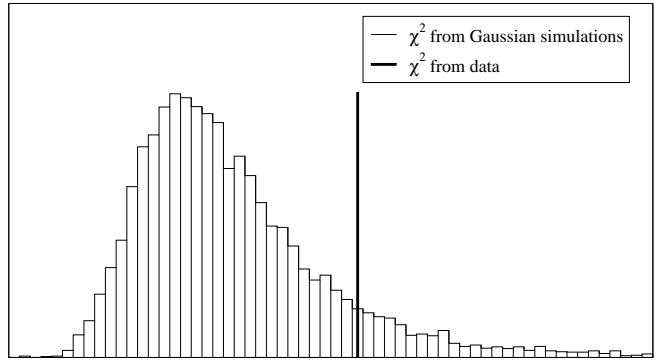}}
\caption{An example of a table with $\chi^2$ obtained from Gaussian simulations (histogram) and a $\chi^2$ from 
real data (line). The fraction of simulated $\chi^2$ above the $\chi^2$ from real data describes its deviation from the Gaussian mean.
}
\label{fig:chisq}
\end{figure}

We now present a method for estimating how much a CMB 
data set deviates from Gaussianity, based on standard statistical tests.
We calculate the covariance matrix $\mathbf C$ from the fraction of hills, lakes and saddles for
$40 000$ Gaussian simulations. We implement a standard $\chi^2$-test on the form
\begin{equation}
\chi^2 = \mathbf{d^T C^{-1} d},
\end{equation}
where $\mathbf d = X_{T_t} - \langle X \rangle_{T_t}$ and $X_{T_t}$ is a hill, lake or saddle density at threshold $T_t$. We then
estimate the $\chi^2$ on $10 000$ simulations, and 
obtain a histogram of the distribution, as presented in 
Figure \ref{fig:chisq}. When performing the analysis on WMAP data, we
first calculate the $\chi^2$ and then compare it to the
pre-calculated $\chi^2$ distribution. We then count the percentage of the
$\chi^2$s that are above the experimental, i.e. if $32 \%$ of the simulated $\chi^2$ values 
are larger than the $\chi^2$ from experimental data this corresponds to $1 \sigma$ deviation
from the Gaussian expectation. A value of $5 \%$ would be consistent with a
$2 \sigma$ deviation from the Gaussianity.

As an alternative method we also use a diagonal covariance matrix for
calculating the $\chi^2$ values, thus ignoring correlations between
threshold levels. This is the method applied in \cite{hansen:2004b}.

\section{Data}
We consider the publicly available 5-year WMAP
data \citep{bennett:2003, hinshaw:2007} that can be obtained from the LAMBDA\footnote{http://lambda.gsfc.nasa.gov/}
site. We primarily perform the analyses on a combined V and W frequency band (94
and 61 GHz) in order to keep noise and foregrounds to a minimum. In addition to
the 5-yr WMAP data, we also simulate $50 000$ Gaussian CMB maps based on
the best-fit angular powerspectrum from the 5-year WMAP release. We
convolve the simulated maps with the combined V and W instrumental
beam before adding Gaussian noise. All CMB maps considered in this paper
have Healpix resolution $N_{\textrm{side}}=512$.

We focus on several different divisions of the sky. First, we
consider the full-sky WMAP data including the extended KQ85 and the KQ75 mask from Section \ref{sec:extended_mask}. We
then investigate the northern and southern galactic hemispheres
individually, using only the KQ85 mask. Due to the alignment of the WMAP satellite with the
ecliptic plane, we also include an analysis of the northern and
southern ecliptic hemispheres. We then perform the full-sky analysis
on a co-added Q-band data set, a co-added V-band set and a co-added 
W-band set. Finally, we re-analyze the co-added V+W data sets on hemispheres centered
around all the pixel centers on a map with $N_{\textrm{side}}=2$. From these results, we obtain
maps with preferred directions depicting the amount of deviation from Gaussianity on the sky. 

\section{Results}

Here we present the results, first with a full covariance matrix and then without.

\subsection{Results with full covariance matrix}

\label{sec:results}
\begin{figure*}
\mbox{\epsfig{file=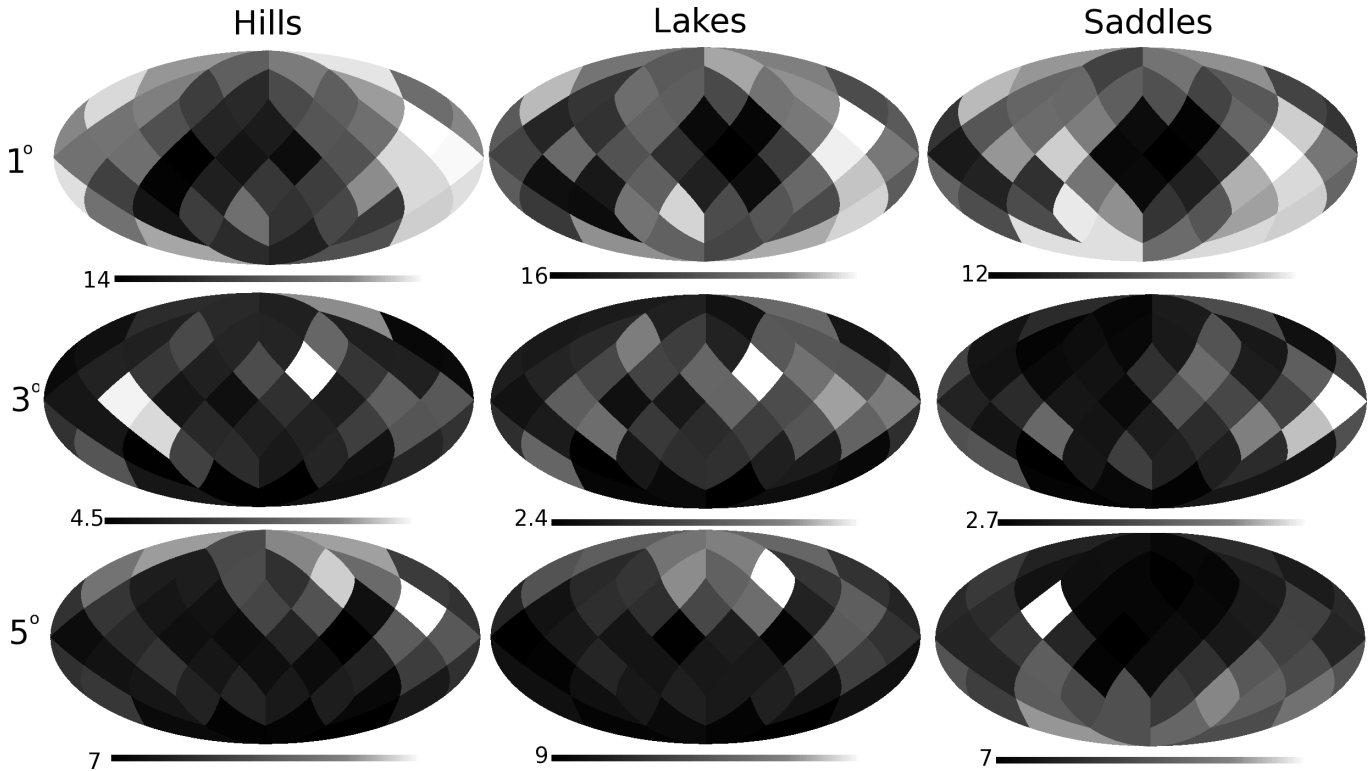, width=\linewidth,clip=}}
\caption{The results from an non-Gaussian analysis of hemispheres centered around pixels on a map with 
$N_{\textrm{side}}=2$. Darker pixels correspond to higher deviations from Gaussianity on a hemisphere centered around
this pixel.  The values of the pixels indicate the percentage of simulations with a higher $\chi^2$.
Whiter pixels correspond to $< 1\sigma$ levels.
}
\label{fig:directions}
\end{figure*}
The results from the V+W full sky and  north/south galactic/ecliptic analyses are presented in Table \ref{tab:results1}.
The most interesting result from Table \ref{tab:results1} is the $\sim1\%$ deviation from Gaussianity in the full-sky at $3^\circ$ when counting saddles. 
A general trait throughout the analysis is that
scales above and below $3^\circ$ have deviation from Gaussianity
no more than $2\sigma$. We continue by analyzing whether the $1\%$ detection on the full-sky can be observed
in the various WMAP frequency bands alone. We therefore perform the same full-sky analysis on the 
Q, V and W frequency bands. The results are shown in Table \ref{tab:results2}, and are consistent with the
combined V+W full-sky analysis, where a $2 \sigma$ deviation from Gaussianity is visible in all band for saddles at $3^\circ$. 

Motivated by the asymmetry found in \cite{hansen:2004b}, we perform
the analysis on all hemispheres centered around pixels on a map
with $N_{\textrm{side}}=2$ using the co-added V+W data set. This
analysis used the extended KQ85 mask. The results are shown in Figure
\ref{fig:directions}, where darker pixels represent larger
deviations from Gaussianity on a hemisphere centered around the
pixel. Here, whiter pixels correspond to $< 1\sigma$ levels. These
results agree with the full-sky analysis and with
the ecliptic/galactic hemispherical analyses, and show that there is
no apparent preferred direction in either of the data sets.

Finally, we perform a $\chi^2$ test of the combined results.  
We construct a data vector containing all
fractions of hills and lakes for all thresholds and smoothing
scales ($1^\circ$, $3^\circ$ and $5^\circ$). The covariance
matrix including all correlations between hills and lakes and between different scales, 
is calculated from Gaussian simulations. We estimate a $\chi^2$ value 
for both WMAP data and simulations. The final comparison
shows that $80 \%$ of the simulations have a higher $\chi^2$ value
than the simulations for the combined V + W channel. In other words, when using the full covariance
matrix, there is full consistency with the Gaussian hypothesis.

\begin{deluxetable}{lcccl}
\tablewidth{0pt}
\tablecaption{Summary of deviations from Gaussianity in the combined V+W data} 
\tablecomments{The values correspond to the fraction of $\chi^2$ from isotropic simulations
that have a higher $\chi^2$ than the analysis of the data set. A low percentage would thus indicate a non-Gaussianity.} 
\tablecolumns{4}
\tablehead{ Area/Scale & \hspace{5mm} $60'$ \hspace{5mm} & \hspace{5mm}$180'$\hspace{5mm} & \hspace{5mm}$300'$}
\startdata
\cutinhead{Full sky + KQ85 mask}
hills     & $27 \%$ & $6.7 \%$ & $3.87 \%$ \\
saddles   & $69 \%$ & $1.13\%$& $11 \%$ \\
lakes     & $46 \%$ & $15 \%$ & $6.5 \%$ \\
\cutinhead{Full sky + KQ75 mask}
hills     & $48 \%$ & $10 \%$ & $14 \%$ \\
saddles   & $85 \%$ & $1.33 \%$& $7 \%$ \\
lakes     & $92 \%$ & $14 \%$ & $16 \%$ \\
\cutinhead{Northern galactic hemisphere}
hills     & $86 \%$ & $30 \%$ & $46 \%$ \\
saddles   & $75 \%$ & $13 \%$ & $25 \%$ \\
lakes     & $73 \%$ & $18 \%$ & $38 \%$ \\
\cutinhead{Southern galactic hemisphere}
hills     & $39 \%$ & $5 \%$ & $10 \%$ \\
saddles   & $78 \%$ & $5 \%$ & $54 \%$ \\
lakes     & $53 \%$ & $5 \%$ & $12 \%$ \\
\cutinhead{Northern ecliptic hemisphere}
hills     & $64 \%$ & $28 \%$ & $15 \%$ \\
saddles   & $83 \%$ & $5 \%$ & $20 \%$ \\
lakes     & $69 \%$  & $45 \%$ & $25 \%$ \\
\cutinhead{Southern ecliptic hemisphere}
hills     & $73 \%$ & $13 \%$ & $13 \%$ \\
saddles   & $75 \%$ & $15 \%$ & $54 \%$ \\
lakes     & $62 \%$ & $14 \%$ & $17 \%$ \\
\enddata
\label{tab:results1}
\end{deluxetable}

\begin{deluxetable}{lcccl}
\tablewidth{0pt}
\tablecaption{Summary of deviations from Gaussianity in the combined
  V+W data using a strictly diagonal covariance matrix} 
\tablecomments{The values correspond to the fraction of $\chi^2$ from isotropic simulations
that have a higher $\chi^2$ than the analysis of the data set. A low percentage would thus indicate a non-Gaussianity.} 
\tablecolumns{4}
\tablehead{ Area/Scale & \hspace{5mm} $60'$ \hspace{5mm} & \hspace{5mm}$180'$\hspace{5mm} & \hspace{5mm}$300'$}
\startdata
\cutinhead{Full sky + KQ85 mask}
hills     & $12 \%$ & $0.95 \%$ & $0.68 \%$\\
saddles   & $11 \%$ & $0.034\%$& $10 \%$ \\
lakes     & $11 \%$ & $0.67 \%$ & $0.64 \%$\\
\cutinhead{Northern ecliptic hemisphere}
hills     & $ 13 \%$ & $ 0.38  \%$ & $ 4.2  \%$ \\
saddles   & $ 11 \%$ & $ 0.34  \%$ & $ 11  \%$ \\
lakes     & $ 3.8 \%$ & $ 0.67  \%$ & $ 4.2  \%$ \\
\cutinhead{Southern ecliptic hemisphere} 
hills     & $ 39  \%$ & $ 30  \%$ & $ 15  \%$ \\
saddles   & $ 78  \%$ & $ 4.0  \%$ & $ 87  \%$ \\
lakes     & $ 29  \%$ & $ 3.4  \%$ & $ 12  \%$ \\
\enddata
\label{tab:resultsdiagonal}
\end{deluxetable}

\begin{deluxetable}{lccc}
\tablewidth{0pt}
\tablecaption{Summary of deviations from Gaussianity in full-sky combined bands} 
\tablecomments{The values correspond to the fraction of $\chi^2$ from isotropic simulations
that have a higher $\chi^2$ than the analysis of the data set. A low percentage would thus indicate a non-Gaussianity.} 
\tablecolumns{4}
\tablehead{ Area/Scale & \hspace{5mm} $60'$ \hspace{5mm} & \hspace{5mm}$180'$\hspace{5mm} & \hspace{5mm}$300'$}
\startdata
\cutinhead{Full-sky combined W-band}
hills     & $20 \%$ & $7.5 \%$      & $5.9 \%$  \\
saddles   & $66 \%$ & $2.8 \%$      & $24 \%$ \\
lakes     & $37 \%$ & $27 \%$      & $5 \%$  \\
\cutinhead{Full-sky combined V-band}
hills     & $24 \%$ & $11 \%$      & $8.2 \%$  \\
saddles   & $39 \%$ & $1.61 \%$  & $13 \%$  \\
lakes     & $65 \%$ & $27 \%$      & $7 \%$  \\
\cutinhead{Full-sky combined Q-band}
hills     & $15 \%$ & $6.6 \%$       & $3.13 \%$  \\
saddles   & $76 \%$ & $1.73 \%$       & $8.9 \%$    \\
lakes     & $34 \%$ & $25 \%$      & $5 \%$  \\

\enddata
\label{tab:results2}
\end{deluxetable}

\subsection{Comparison with earlier work} 
\cite{hansen:2004b} found very little evidence for deviation from
Gaussianity in the 1-year WMAP data when analyzing full-sky CMB maps
using the Kp2 galaxy mask. This is in good agreement with the findings in
this paper, where we have shown there is at most a $2\sigma$ deviation from
Gaussianity in the saddles at $3^\circ$ in the combined V+W data. We
stress that while the number of simulated Gaussian maps used for the
$\chi^2$ test by \cite{hansen:2004b} was 512, we have used 50
000. Where \cite{hansen:2004b} employed a covariance matrix that was
strictly diagonal, we have operated with a full covariance matrix
including correlations between different thresholds.  In addition, the
1-year WMAP data are in general more contaminated by instrumental
noise than the 5-year data.

When using the full covariance matrix, the directional analysis show
that there isn't any preferred direction for deviations from
Gaussianity. Figure \ref{fig:directions} shows that the directions for
various scales seem to be randomly scattered, and the significance is
very low.  In fact, only at $3^\circ$ there seems to be a $2 \sigma$
hint, but any exact directions are non-existent.  This is in
disagreement with the results from \cite{hansen:2004b}, who claimed
that there exists a maximum of non-Gaussianity on hemispheres centered
at the ecliptic poles. 

\begin{figure*}
\mbox{\epsfig{file=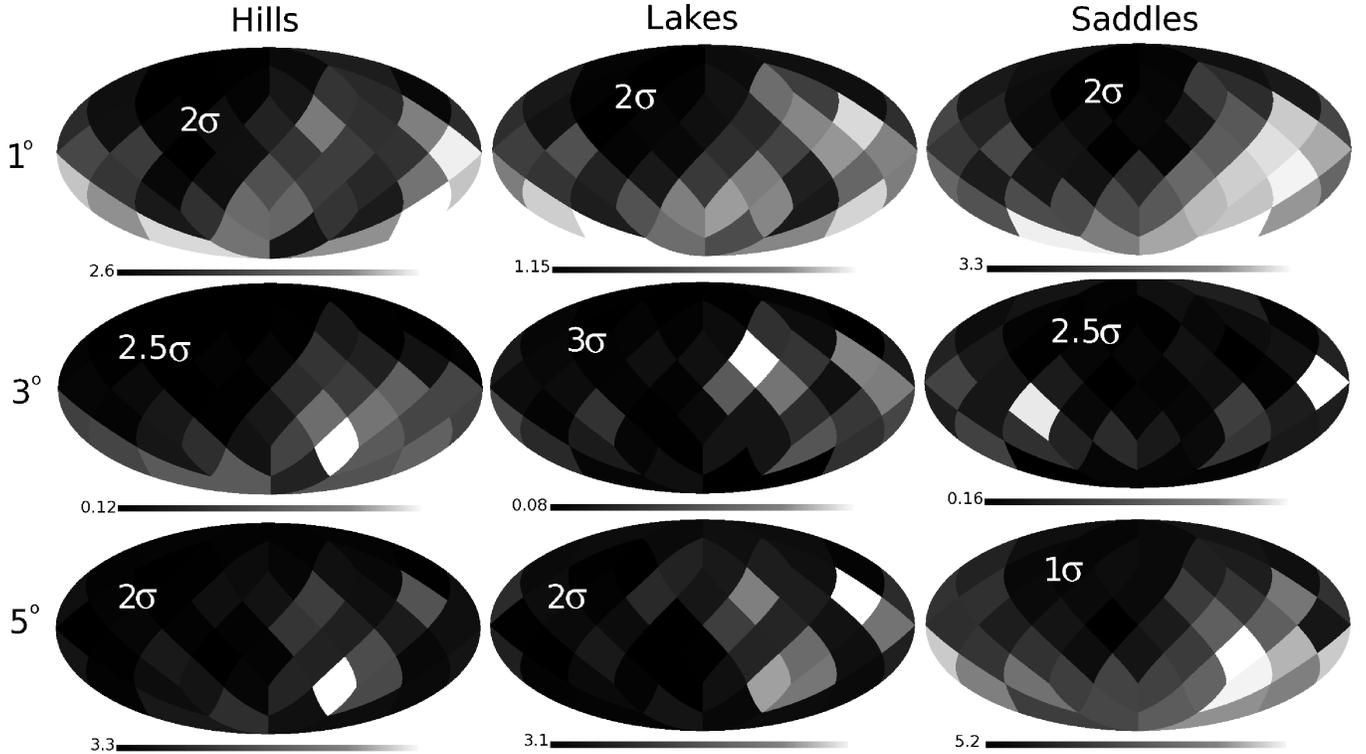, width=\linewidth,clip=}}
\caption{The results from a non-Gaussian analysis using a diagonal
  covariance matrix, of hemispheres centered around pixels on a map
  with $N_{\textrm{side}}=2$. Darker pixels correspond to higher
  deviations from Gaussianity on a hemisphere centered around this
  pixel. The values of the pixels indicate the percentage of
  simulations with a higher $\chi^2$.  Whiter pixels correspond to $<
  1\sigma$ levels.}
\label{fig:directions_old}
\end{figure*}

However, when we ignore correlations and use a strictly diagonal
covariance matrix, the results are more in agreement with
\cite{hansen:2004b}. These results are presented in Table
\ref{tab:resultsdiagonal}, where we note several deviations at the
less than $1\%$ level. Also, when performing the directional analysis,
we find that the non-Gaussian signal has a clear maximum close to (but
not directly on) the ecliptic north-pole, as seen in figure
\ref{fig:directions_old} and Table \ref{tab:resultsdiagonal}.  This is
in agreement with the direction described by \cite{hansen:2004b}. As
noted above, here we use a larger extended mask than in
\cite{hansen:2004b}.  We have also tested with the smaller extended
mask and find results simiar to the ones presented in Table
\ref{tab:resultsdiagonal}.  We conclude that by using a diagonal
approximation to the correlation matrix, we still find
non-Gaussianities and asymmetries over several scales and with
different masks, including the full covariance matrix lowers the
effect of the non-Gaussian signatures and the hemispherical
anisotropy.

\subsection{Tests on weak lensing simulations}
In order to investigate whether the deviations may be a result of weak
lensing in the CMB, we performed another test. We used the freely
available LensPix\footnote{http://cosmologist.info/lenspix/} code
(\cite{lenspix}) to simulate $300$ Gaussian CMB maps, and $300$ CMB
maps with weak lensing. For the simulations we used the best-fit power
spectrum and lensing potential provided by the WMAP team. The
procedure of counting hills, lakes and saddles as described above was
applied to all the simulations. From the Gaussian simulations we found
mean values and standard deviations of hill, lake and saddle fractions
at different smoothing scales. We then compared these with the
results from the simulations with weak lensing. We found no
significant deviation in the local curvature of the simulations with
lensing. We therfore conclude that the deviations from non-Gaussianity
described in this paper and \cite{hansen:2004b} are not caused by weak
lensing in the CMB.

\section{Conclusions}
\label{sec:conclusions}
We have developed an independent framework for estimating deviations
from Gaussianity in CMB data based on the methods established by
\cite{dore:2003} and \cite{hansen:2004b}. The methods used are
model-independent, and do not share any obvious connections with
non-Gaussianity frameworks of known physical origin. By counting the
fraction of lakes, hills and saddles in simulated Gaussian maps while
increasing the temperature threshold, we have built a distribution for
what is expected for Gaussian maps. We then compared experimental data
to this distribution, determining the deviation from the Gaussian
assumption.  We then considered a combined V + W full-sky data set
with the extended KQ85 and KQ75 mask, and found evidence of a
$\sim1\%$ deviation from Gaussianity on scales around $3^\circ$. We
also analyzed other scales as well as the north and south
galactic/ecliptic hemispheres seperately, but discovered no deviation
from Gaussianity greater than $2 \sigma$. We continued by performing
an analysis on each of the hemispheres centered around a pixel on a
Healpix map with $N_{\textrm{side}}=2$ using the combined V+W data. We
produced directional maps for hills, lakes, saddles using the three
scales, and found no evidence for a preferred direction in either of
the maps. Finally, we calculated the combined $\chi^2$ from all our
results, which resulted in an overall agreement with Gaussianity. We
conclude that there is no significant evidence for non-Gaussianities
or asymmetries in the WMAP data based on this test.

However, in \cite{hansen:2004b}, it was found that the northern
ecliptic hemisphere was non-Gaussian based on similar local curvature
measurements. There was however one large difference in the method
used: In that work, a diagonal approximation to the covariance matrix
was applied. Repeating our analysis with a diagonal covariance matrix
we obtain similar results as \cite{hansen:2004b}. Taking into account
correlations between thresholds, the non-Gaussianity disappears.  We
have compared Gaussian CMB simulations to CMB simulations with weak
lensing to see whether the hill, lake or saddle densities are
different in the two sets of simulations. No significant differences
were found. We therefore conclude that weak lensing may not cause the
non-gaussianity found when using a diagonal correlation matrix. It is
still unclear what causes the detection of non-Gaussianity when a
diagonal correlation matrix is used. Even though the $\chi^2$ test is
not optimal when correlations are ignored, we are still comparing the
data to simulations for which an identical procedure (i.e. diagonal
approximation to the covariance matrix) has been applied. If the best
fit model estimated from the data is correct, one would expect
simulations based on this model to have the same statistical
properties as the data, no matter which statistical test is
performed. There is thus still a discrepancy between data and
simulations based on the model which best fits the data. Whether this
discrepancy is a statistical fluke or may arise from systematic
errors/cosmology and whether it is related to other asymmetries is
still unclear.

\begin{acknowledgements}
  FKH and NEG acknowledge support from the research council of Norway. We acknowledge use of the
  HEALPix\footnote{http://healpix.jpl.nasa.gov} software
  \citep{gorski:2005} and analysis package for deriving the results in
  this paper. We acknowledge the use of the Legacy Archive for
  Microwave Background Data Analysis (LAMBDA). Support for LAMBDA is
  provided by the NASA Office of Space Science.
\end{acknowledgements}

\end{document}